\documentclass[12pt]{spieman}  
\usepackage{amsmath,amsfonts,amssymb}

\usepackage{graphicx}
\usepackage{url}
\usepackage{hyperref}
\hypersetup{
    colorlinks=true,
    linkcolor=red,
    filecolor=magenta,      
    urlcolor=magenta,
    citecolor=blue,
}
\usepackage{txfonts}
\usepackage{setspace}
\usepackage{tocloft}

\title{A prototype for pulsar observations at low radio frequencies using log periodic dipole antennas}

\author[a,*]{Kshitij S. Bane}
\author[b]{Indrajit V. Barve}
\author[b]{G. V. S. Gireesh}
\author[a]{C. Kathiravan}
\author[a]{R. Ramesh}
\affil[a]{Indian Institute of Astrophysics, 2nd Block, Koramangala, Bangalore 560034, India}
\affil[b]{Radio Astronomy Field Station, Indian Institute of Astrophysics, Gauribidanur 561210, India}

\cftpagenumbersoff{figure}
\cftpagenumbersoff{table} 
\begin{document} 
\maketitle

\begin{abstract}
A prototype for dedicated observations of pulsars and other astrophysical transients in the frequency range 50\,-\,80\,MHz has been recently commissioned at the Gauribidanur radio observatory near Bangalore in India. The antenna set-up, analog \& digital receiver systems, and the initial observations are presented.
\end{abstract}


{\noindent \footnotesize\textbf{*}\linkable{kshitij.sb@iiap.res.in}}

\begin{spacing}{2}   

\section{Introduction}
\label{sect:intro}  
A pulsar is a neutron star, i.e. the ultra-dense core that remains after a massive star undergoes a supernova explosion. It spins at very rapid rates ranging from once in a few seconds to as much as ${\sim}$700 times per second. 
The strong magnetic fields (${\sim}10^{8}{-}10^{14}$\,G) created when the neutron star was formed and the star's rapid rotation results in a
magnetosphere.
Due to misalignment of the magnetic and rotation axes of the pulsar, the radio emission 
comes out in two beams, one from each pole
of the magnetosphere. The observed emission is coherent, but the mechanism behind the coherence is still being debated. These rotating beams of radiation could be noticed whenever they
intersect our line of sight to the pulsar, much like a lighthouse on the sea-shore. Each rotation of the pulsar thus produces a narrow pulse of radiation.
The first pulsar was observed at a low frequency of ${\approx}$\,81.5\,MHz\cite{Hewish68}. But, much of our present knowledge about pulsars have come primarily from radio observations at frequencies 
$>$300\,MHz.
The primary reason for this is that the emission from the pulsars at low frequencies are more affected by dispersion and scattering during their propagation through the interstellar medium (ISM) as compared to higher frequencies. While the delay in the arrival time of the pulses due to the dispersive properties of the ISM varies as 
$\nu^{-2}$ (where $\nu$ is the frequency of observation), the interstellar scattering changes as ${\sim}\nu^{-4}$ (see e.g. \citenum{Stovall15}). The latter leads to broadening of the pulses as a result of which they overlap\cite{Bhat04,Geyer17}. 
It then becomes difficult to distinguish the maxima and minima in a pulse. The periodic nature of the signal is either subdued or nearly lost when the pulse broadening time becomes nearly comparable to the spin period of the pulsar (see e.g. \citenum{Gupta16}).
The other reasons for the limited observations of pulsars at frequencies 
$\lesssim$100\,MHz are the comparatively intense synchrotron emission from 
the Galactic background, 
and the radio frequency interference (RFI). 

Pulsar observations at frequencies
$\lesssim$100\,MHz are necessary to understand the emission mechanism and the characteristics of the pulse profile as a function of frequency which are still being examined. For e.g., the radius-to-frequency mapping (RFM) effect in pulsars are most prominent at lower frequencies\cite{Cordes78,Thorsett91}. Many pulsars exhibit a turn-over in the spectrum close to 
${\approx}$100\,MHz\cite{Malofeev94,Bilous16}. Therefore the extension of their spectra towards lower radio frequencies is naturally of great interest. This is especially so for those pulsars whose intensity is increasing with decreasing frequency in the known part of their spectra. An understanding of the propagation of the low frequency radio waves in the ISM is also important. The $\nu^{-4}$ dependence of scattering in the ISM makes the dispersion effects more stronger at lower frequencies and hence it is easier to carry out better measurements of the dispersion measure (DM). 
Observations of pulsars at frequencies 
$\lesssim$100\,MHz are currently carried out with LOFAR\cite{Stappers11,Bilous20}, UTR-2 telescope\cite{Zakharenko13}, MWA\cite{Tingay13}, LWA\cite{Stovall15}, and EDA\cite{Wayth17,Sokolowski21} 
in the time sharing mode. Exploration of transient phenomena in the Universe is an exciting and rapidly growing area of radio astronomy. Considering these, we have set-up a 
radio telescope in the Gauribidanur observatory\cite{Sastry95,Ramesh11} 
(Longitude:\,$77.4^{\circ}$\,E; Latitude:\,$13.6^{\circ}$\,N) for dedicated observations of pulsars in the frequency range 50\,-\,80\,MHz. RFI in Gauribidanur is also minimal\cite{Monstein07,Hariharan16,Mugundhan18}.
We propose to observe and understand the characteristics of the known pulsars and the periodic Fast Radio Bursts (FRBs)\cite{Thornton13,Fedorova19,Pleunis21,Ines21} at low frequencies.
The observing facility described in the work is a dedicated instrument with wide sky coverage, large bandwidth, and high time resolution (see Sections 2 and 3) all of which are important for observations of fast transients (see e.g. \citenum{Gupta16}).
In the rapidly developing field of study of the transient sky,
FRBs are perhaps the most exciting objects of scrutiny at
present. So there is potential for new observations\cite{Maan15}.

\section{The Antenna and Front-end Analog Receiver System}
\label{sect:antenna}  

The basic signal reception element used in the present case is a Log Periodic Dipole Antenna (LPDA). It is a type of broadband directional antenna whose characteristics are nearly constant over its entire operating frequency range\cite{Rumsey57}. We have set-up an array of 16 LPDAs on a north-south baseline with a spacing of 5\,m between the adjacent antennas. They are combined as two groups with 8 LPDAs in each group (see Figure 1). The LPDAs used in the array are designed according to the formulations reported in the literature\cite{Carrell61}. The characteristic impedance of the LPDA is ${\approx}$50\,${\Omega}$. 
Its half-power beam widths (HPBW) as per calculations are 
${\approx}80^{\circ}$ (E-plane) and 
${\approx}110^{\circ}$(H-plane). All the 16 LPDAs have been mounted with their 
H-plane in the east-west direction. 
The effective collecting area and gain of each LPDA is ${\approx}$0.4${\lambda}^{2}$ (where ${\lambda}$ is the wavelength corresponding to the observing frequency) and ${\approx}$6.5\,dBi (with respect to an isotropic radiator), respectively. The Voltage Standing Wave Ratio (VSWR) is $<$2 over the frequency range ${\approx}$40\,-\,440\,MHz\cite{Kishore14}. The aforementioned effective collecting area and the HPBW of the LPDA are larger compared to that of the other similar frequency independent receiving elements like the inverted V-shaped dipoles and bowtie antennas, respectively, used elsewhere. Since the LPDAs are arranged on a north-south baseline, the combined response pattern of the array (16 LPDAs) is 
${\approx}110^{\circ}$ in the east-west (hour angle) direction and 
${\approx}3^{\circ}$ in the north-south (declination) direction for observations near the zenith at a typical frequency like 65\,MHz. It is a fan-beam with the above resolution in declination. The width of the east-west response pattern is nearly independent of frequency. Being very wide, it helps to observe a radio source
continuously for $\gtrsim$7\,h.
A RG58U coaxial cable connected to the feedpoint near the top of the LPDA is used to transmit the radio frequency (RF) signal incident on the LPDA to the input of a low-pass filter with 3\,dB cut-off at ${\approx}$85\,MHz,  followed by a high-pass filter with 3\,dB cut-off at ${\approx}$35\,MHz and a wideband amplifier with an uniform gain ${\approx}$30\,dB in the frequency range 35-85\,MHz. The two filters and the amplifier are kept near the base of the LPDA to minimize the length of the RG58U cable and hence the transmission loss. The high- and low-pass filters help to attenuate the unwanted signal at frequencies ${\lesssim}$30\,MHz and ${\gtrsim}$85\,MHz.
The filtered and amplified signal from each LPDA is then passed via a cable delay unit (see the following paragraph). Subsequently, they are combined using two sets of 8-way power combiner followed by again a high-pass filter, low-pass filter and wideband amplifier as mentioned above. 

\begin{figure}[!t]
\begin{center}
\begin{tabular}{c}
\includegraphics[width=15cm]{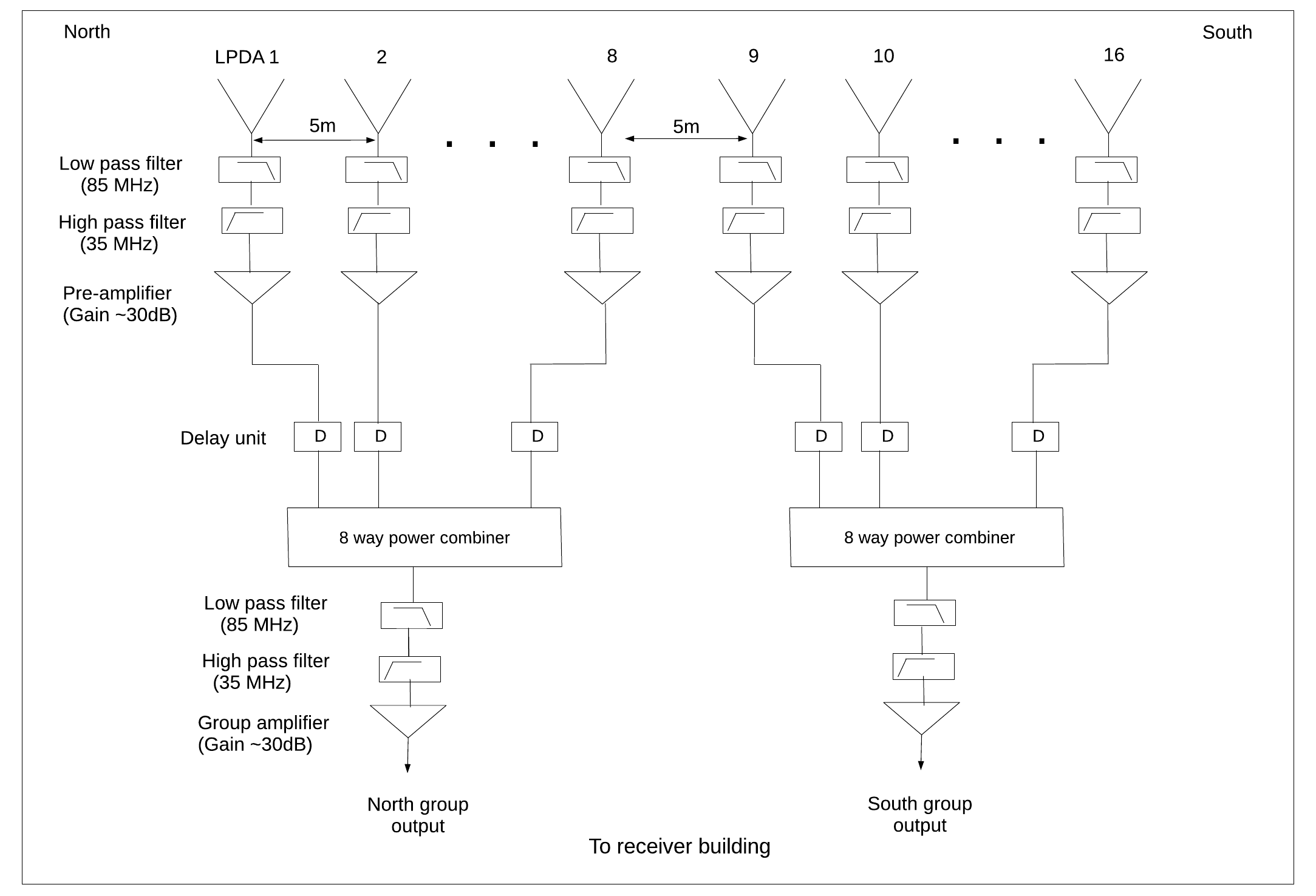}
\end{tabular}
\end{center}
\caption 
{\label{fig:figure1}
The array configuration and the front-end signal path.} 
\end{figure} 

As the array is oriented in the north-south direction, RF signal coming from an astronomical source whose declination is different from the local zenith (${\approx}14^{\circ}$\,N) will be incident on each LPDA in the array at different instances of time. So, in order to coherently combine the signal from the different LPDAs, there is provision to include cable of appropriate length in the signal path from each LPDA in the array to compensate the aforesaid time delay.  This allows the response pattern (`beam') of the each group of eight LPDAs to be `phased' towards any desired declination in the range 
${\approx}{-26^{\circ}}$\,S to ${\approx}{+54^{\circ}}$\,N. The limits are due to the HPBW of the LPDA along its E-plane and the local latitude mentioned earlier. The step interval in the cable delay unit corresponds to an angle of 
${\approx}3^{\circ}$, and
the maximum angle to which the  `beam' of the group of eight LPDAs could be `phased' in the north-south direction is ${\approx}45^{\circ}$.
Note that the aforementioned minimum step interval in the delay unit results in a phasing error
of ${\pm}1.5^{\circ}$ (same as that for the `beam' of the array in declination mentioned in the previous paragraph). It corresponds to a delay error of ${\pm}$3\,nsec for the total length of 35\,m for the group of eight LPDAs (see Figure 1). We neglect this since it is small in general and also lesser than the minimum delay of ${\approx}$11\,nsec that could be applied digitally in the present case (see Section 3).
The combined output from each group of antennas are then transmitted to the receiver building (located 
${\approx}$300\,m away) via low-loss coaxial cables buried ${\approx}$1\,m below the ground level to minimize possible diurnal variations in the characteristics of the cable. In order to compensate for the transmission losses during propagation, the RF signal from the two antenna groups are amplified in the receiver building. The output of the amplifiers are passed through 50\,-\,80\,MHz bandpass filters to attenuate any spurious pick-up outside the band. Then the signal is fed to the input of the digital receiver.

\begin{figure}[!t]
\begin{center}
\begin{tabular}{c}
\includegraphics[width=15cm]{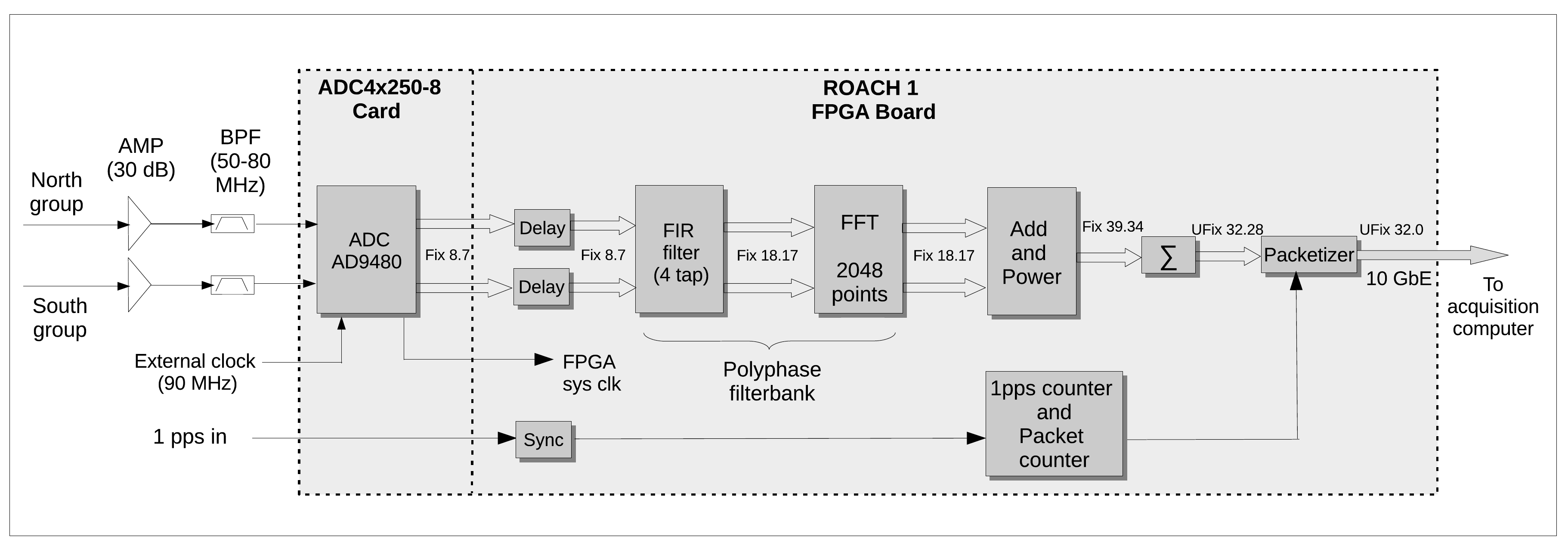}
\end{tabular}
\end{center}
\caption 
{\label{fig:figure2}
Schematic of the digital receiver system.} 
\end{figure} 

\section{The Back-end Digital Receiver System}
\label{sect:digital}  

We have used a Field Programmable Gate Array 
(FPGA)\footnote{\url{https://www.xilinx.com/products/silicon-devices/fpga/what-is-an-fpga.html}} based digital spectrometer which converts the input voltage signal to its power spectrum using the Fast Fourier Transformation (FFT) technique. The spectrometer is implemented on Reconfigurable Open Architecture Computing Hardware (ROACH) from the Collaboration for Astronomy Signal Processing and Electronics Research (CASPER). The ROACH board hardware has Xilinx Virtex-5 
FPGA\footnote{\url{https://casper.astro.berkeley.edu/wiki/ROACH}}. Figure 2 shows the schematic of the digital receiver system used in the present case. 

The RF signals corresponding to the two antenna groups in Figure 1 are connected to the input of an analog-to-digital converter (ADC). It is a quad-ADC containing four AD9480 
ICs\footnote{\url{https://www.analog.com/en/products/ad9480.html}}. We use two channels of the ADC for the present work. The ADC converts the input voltage to a 8-bit Fixed point number (Fix 8.7) between -1 and +1. Note that a linear relation between the input and
output of the ADC for a large range of the input signal is important for any digital system. To check this in the present case, a broadband noise signal with provision to vary the input power level was connected to the input of the ADC. The output from the ADC were recorded for different levels of the input signal, and then processed offline to get the power spectrum using Fast Fourier Transformation (FFT). Figure 3 shows the results of the above test. The spectral power is linear for the input signal amplitude in the range 
${\approx}$\,-42\,dBm to ${\approx}$\,-4\,dBm. This implies that the measured dynamic range (MDR) of the ADC is ${\approx}$\,38\,dB. The effective number of bits (ENOB)\,=\,MDR/6.02\,=\,6.3 bits\footnote{\url{https://www.allaboutcircuits.com/technical-articles/understanding-the-dynamic-range-specification-of-an-ADC/}}.
Note that for a 8-bit ADC, the expected 
dynamic range (EDR) and ENOB are 
${\approx}$\,6.02${\times}$8\,=\,48\,dB, and ${\approx}$\,7.3 bits\footnote{\url{https://www.analog.com/media/en/technical-documentation/data-sheets/AD9480.pdf}},
respectively. 
But these numbers are valid primarily for continuous wave (CW) signals at specific spot frequencies only.

\begin{figure}[!t]
\begin{center}
\begin{tabular}{c}
\includegraphics[width=15cm]{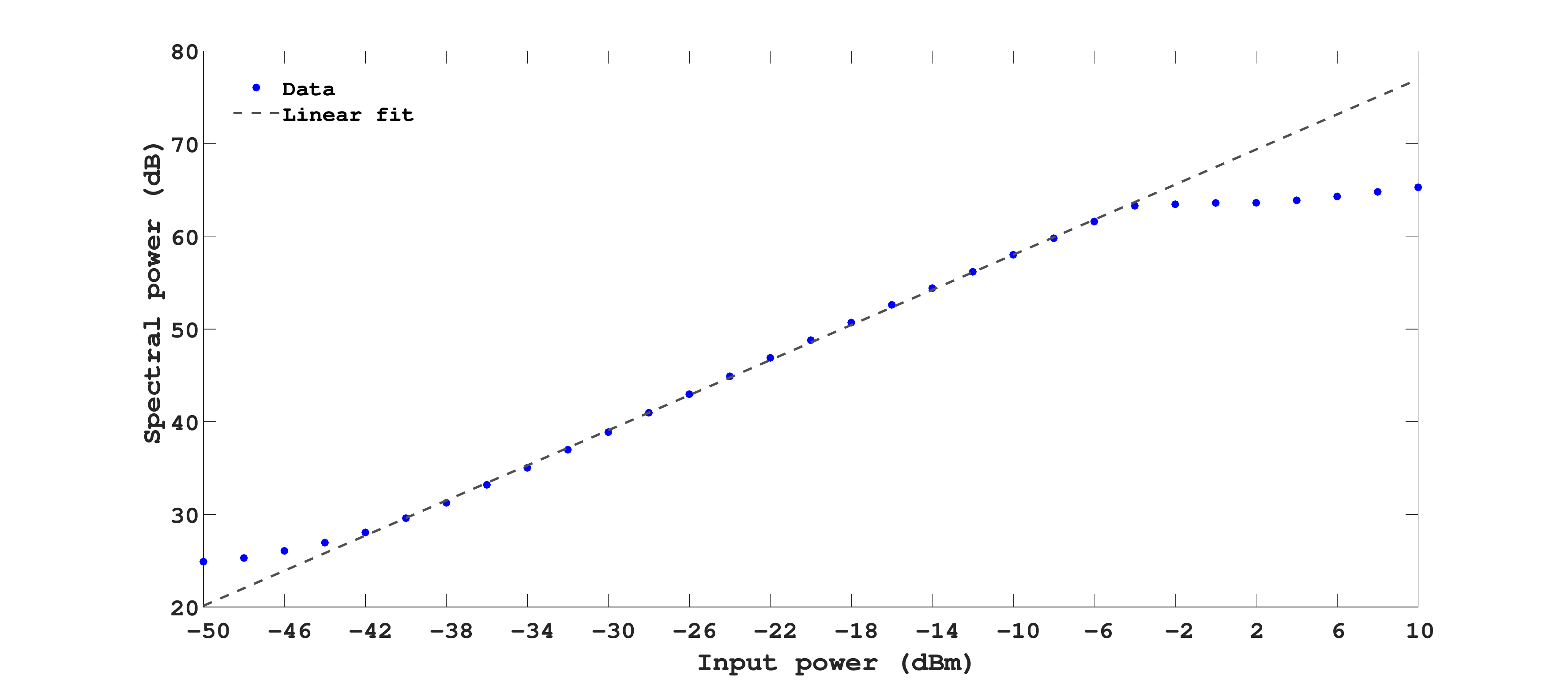}
\end{tabular}
\end{center}
\caption 
{\label{fig:figure3}
Results of the ADC linearity test.} 
\end{figure} 

We use a 90\,MHz clock to sample the bandlimited RF signal of 50-80\,MHz in the present case since the latter is in the 2nd Nyquist zone (45\,-\,90\,MHz) of the aforementioned sampling 
frequency\footnote{\url{https://www.mikrocontroller.net/attachment/341426/Understanding{\_}digital{\_}signal{\_}processing.pdf}}. Depending upon the declination of the source, appropriate delays are applied to the RF signal from either of the two groups in steps of 11.11\,nsec to coherently combine the signals later. The corresponding delay error (i.e. ${\pm}$5.5\,nsec) limits the phasing of the north-south `beam' of the array to ${\pm}2.3^{\circ}$ with respect to the source declination. To obtain the power spectra of the RF voltage signals, the outputs from the delay units are processed using a combination of a finite impulse response (FIR) filter and FFT, together called a `Polyphase filterbank' (PFB). Since the FFT is computed over finite number of samples, its response suffers from spectral leakage. But when a FIR filter with windowing function
is used before the FFT, leakage into the adjacent spectral channels is considerably reduced\cite{Price18}. In our design, a 4-tap FIR filter with Hamming window of 18-bit coefficients followed by 2048-points FFT is used. The latter has 1024 `positive' frequency bins. The total bandwidth sampled in the present case is 45 MHz as mentioned earlier. So, the frequency resolution is 43.945\,kHz. 
To test the spectral leakage, CW signal at different frequencies (64.9\,-\,65.1\,MHz in steps of 1\,kHz) was fed to the input of the digital receiver in succession. 
An inspection of the test results (Figure 4) indicate that the isolation between adjacent frequency bins is ${\approx}$\,-49\,dB. 
The full-width at half-maximum (FWHM) of each frequency bin is 
${\approx}$\,36\,kHz. Note that the latter should have been ${\approx}$\,43.945\,kHz as mentioned above.
But due to the frequency response of the 4-tap PFB, the FWHM of each bin is reduced to  ${\approx}$\,36\,kHz. 
The isolation offered by a 4-Tap PFB (${\approx}$\,50\,dB, see Figure 5) is sufficient for isolating spectral leakage. Note that increasing the number of taps in the PFB will improve the isolation. The main lobe response will be more flatter (see Figure 5 for a comparision between 4-tap and 8-tap PFBs). These are useful to carry out high precision timing studies of pulsars. 
But the possibilities of timing studies at low frequencies as in the present case are restricted since the pulse profiles are scatter broadened. Further, folded / average pulse profiles are only generated on most occassions. So a 4-tap PFB with Hamming window was used.

\begin{figure}[!t]
\begin{center}
\begin{tabular}{c}
\includegraphics[width=15cm]{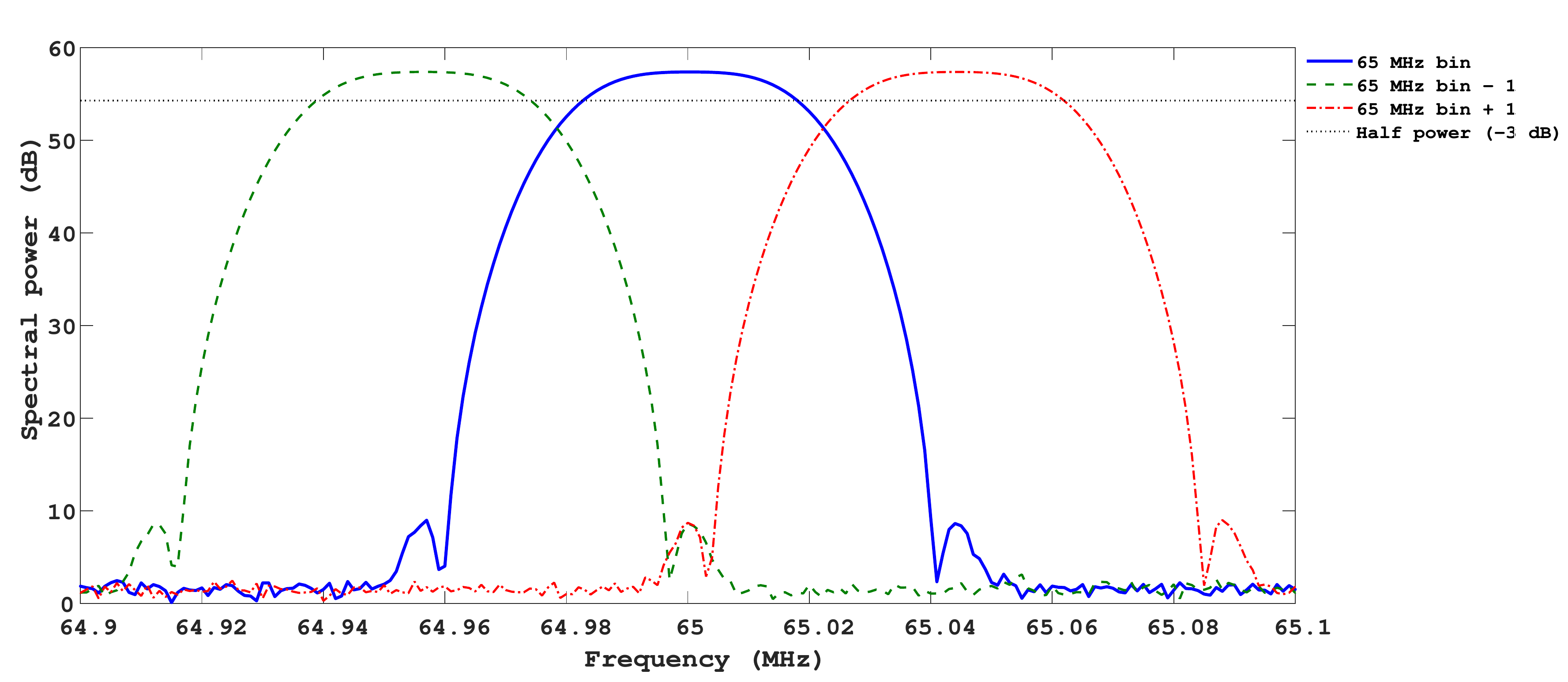}
\end{tabular}
\end{center}
\caption 
{\label{fig:figure4}
Spectral leakage test.} 
\end{figure}  

Note that the pulsar signal gets dispersed during its propogation through the ISM. The higher frequency components in the signal reach the observer earlier compared to the lower frequency components. Therefore the received signal needs to be de-dispersed to reconstruct the pulse. There are two types of de-dispersion methods: coherent and incoherent\cite{Lorimer04}. We use the incoherent de-dispersion technique as it is computationally easier\footnote{\url{http://www.ncra.tifr.res.in/ncra/gmrt/gmrt-users/low-frequency-radio-astronomy/ch17.pdf}}. The frequency resolution mentioned above allows finer correction for the dispersion, and also limits the possible leakage of the narrowband RFI to adjacent spectral channels. The PFB block output is 36-bit wide with 18-bit each for real and imaginary parts. The outputs corresponding to the two antenna groups are combined and squared to get the power spectrum (see Figure 6). The spectral data {\bf are} then integrated for the desired duration. The output at this stage is 39-bit wide. The 10Gb Ethernet interface (see Figure 2) allows 64-bit data to be transmitted\footnote{\url{https://casper-toolflow.readthedocs.io/projects/tutorials/en/latest/tutorials/roach/tut{\_}ten{\_}gbe.html}}. To facilitate the transfer, the aforementioned output is truncated to a  32 bits by leaving 7 bits from the LSB (Least significant bit) and thereby making it easier to packetize the data in such a way that two data points are muxed together. 
Any degradation of the signal due to this truncation is expected to be minimal since the information content in the LSBs correspond primarily to noise fluctuations in the data only.
The integrated output, which is now quantized to a 32-bit unsigned number, 
is packetized with headers and sent to the data acquisition computer over 10\,Gb Ethernet (see Figure 2). Each packet contains a standard UDP header, and custom header. The latter has observation details, 1\,PPS (pulse per second) count, and a packet count. The 1\,PPS  signal is derived from a GPS clock (Trimble Thunderbolt E-GPS Disciplined clock)\footnote{https://timing.trimble.com/wp-content/uploads/thunderbolt-e-gps-disciplined-clock-datasheet.pdf} and given to the FPGA via the ADC card to generate the 1\,PPS count. Packet count is a unique packet number assigned to each packet.

\begin{figure}[!t]
\begin{center}
\begin{tabular}{c}
\includegraphics[width=15cm]{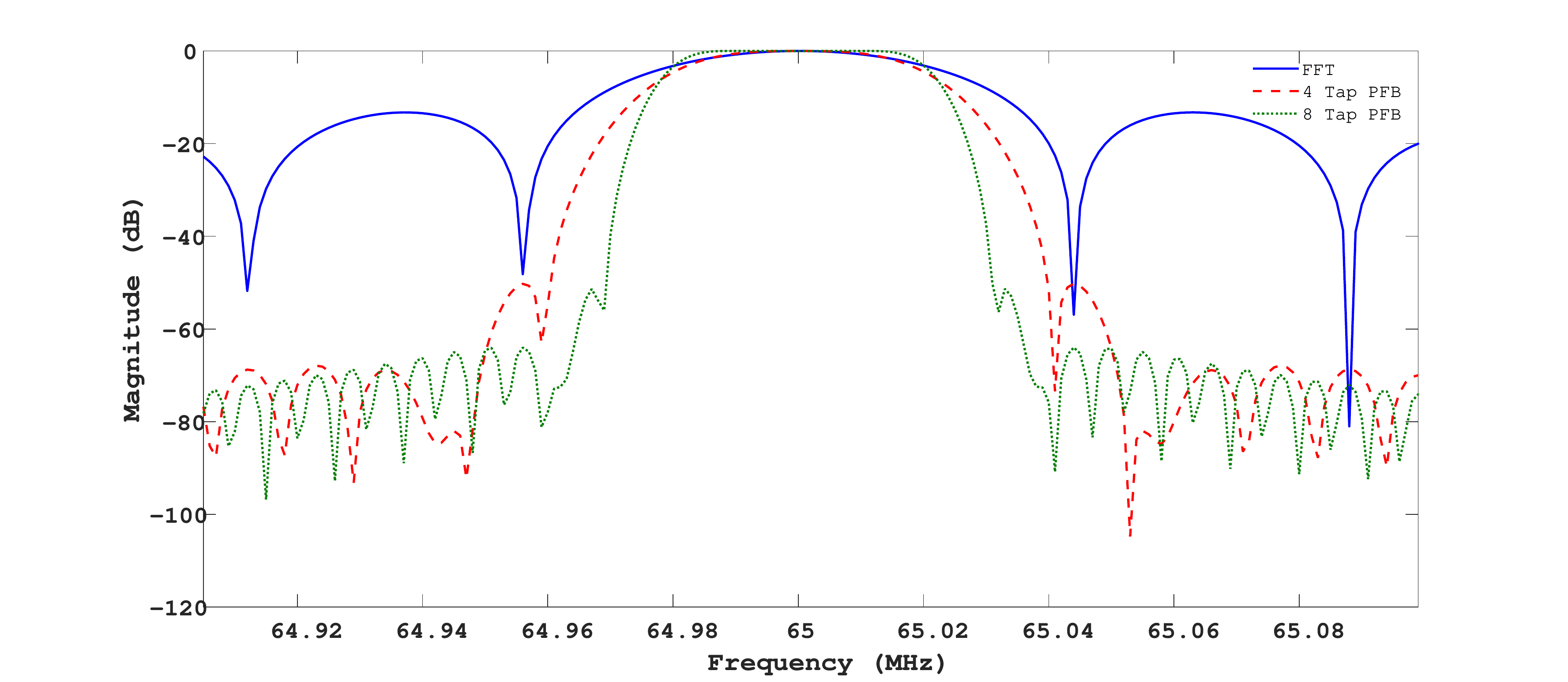}
\end{tabular}
\end{center}
\caption 
{\label{fig:figure5}
Spectral response of 4-tap and 8-tap PFB.} 
\end{figure} 


\section{Data Processing Pipeline}
\label{sect:data}  

\begin{figure}[!t]
\begin{center}
\begin{tabular}{c}
\includegraphics[width=15cm]{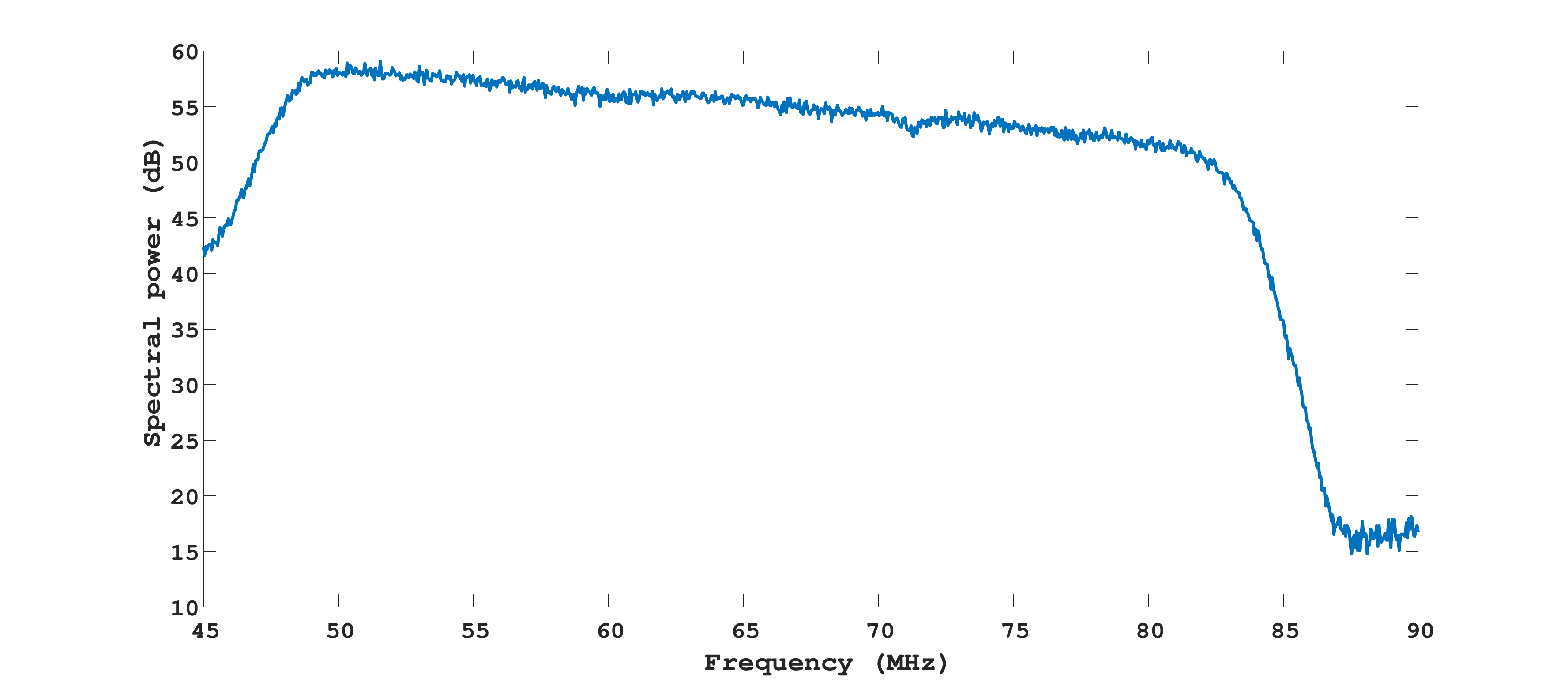}
\end{tabular}
\end{center}
\caption 
{\label{fig:figure6}
A sample power spectrum obtained from observations of the sky background.
The amplitude variation (${\approx}$\,5\,dB) across the 50\,-\,80\,MHz range is due to a combination of the differential attenuation (as a function of frequency) in the coaxial cables used for the RF signal transmission, and the non-thermal spectral nature of the radio emission from the sky background. One could notice that RFI is very minimal as mentioned in Section 1.} 
\end{figure} 

The data acquired are first processed through a MATLAB based quick-look code for a quality check of the observation. The spectral response (see e.g. Figure 6) is plotted and inspected to ensure that the system performance was in order during the observation. The aforesaid code checks for any `packet loss' that might have happened while writing the data to the hard disk of the acquisition computer. The `packet loss' would give rise to gaps in the data stream which in turn affect the arrival times of the pulses leading to  a lower signal-to-noise (S/N) for the observed pulsar. In extreme cases, it may not be possible to detect the pulsar when the data are folded. An unique number is assigned to each packet and the same is mentioned in the file header while writing the data. The packet number is used to assess `packet loss'. If there are any, a new file is created by inserting zeros in the place of the lost packets to maintain continuity in the data stream.
After the above processes, the data are converted to SIGPROC `filterbank' 
format\footnote{\url{http://sigproc.sourceforge.net/sigproc.pdf}} for further analysis using the Pulsar Search and Exploration Toolkit (PRESTO)\cite{Ransom11}. The file is read using the `readfile' tool and the meta-data are examined. RFI mitigation is performed using the `rfifind' tool which searches for prominent RFI in the time-series as well as in the frequency domain,
and creates `mask' files. It also generates diagnostic files containing data statistics, identifies the time-domain statistical issues and marks them as RFI. The RFI `masks' are used in the subsequent stages of data processing. Subsequently, the `prepfold' tool is used to carry out folding and de-dispersion. It searches over a range of pulse periods and DM values to obtain the best fit for the corresponding parameters, i.e. the period and DM of the pulsar present in the observed data. 
Finally, the integrated pulse profile with the highest S/N is generated.


\section{Trial Observations}
\label{sect:result}  

We carried out observations of the sky background in the meridian transit mode to understand the characteristics of the antenna array and the receiver system. Figure 7 shows the observations carried out on 2021 June 30 at a typical frequency like 65\,MHz. The bandwidth and integration time used were ${\approx}$\,30\,MHz and 
${\approx}$\,1\,sec, respectively. The maximum in the observed emission was around 
$\rm{\approx}\,18^{h}$
Local Sidereal Time (LST) as expected (see e.g. \citenum{Kraus66,Kishore15}). 
The duration of the observed profile (at the half maximum level) derived from the least squares fit is
${\approx}$\,9\,h. This corresponds to an angular extent of
${\approx}\,135^{\circ}$. Compared to this, the HPBW of the array pattern (in the east-west direction) is ${\approx}\,110^{\circ}$ (see Section 2). 
Note that the observed deflection in Figure 7 is primarily due to the intense patch of emission extending over the LST range 
${\approx}$\,16\,h\,-\,21\,h in any low frequency all-sky map (see e.g. \citenum{Kraus66,Dwarakanath90}).
A convolution of the aforementioned region 
(angular extent\,${\approx}\,75^{\circ}$) with the array pattern 
(HPBW\,${\approx}\,110^{\circ}$) could give rise to ${\approx}\,135^{\circ}$ width for the observed profile (see e.g. \citenum{Ramesh06,Ramesh20}). 
\begin{figure}[!t]
\begin{center}
\begin{tabular}{c}
\includegraphics[width=15cm]{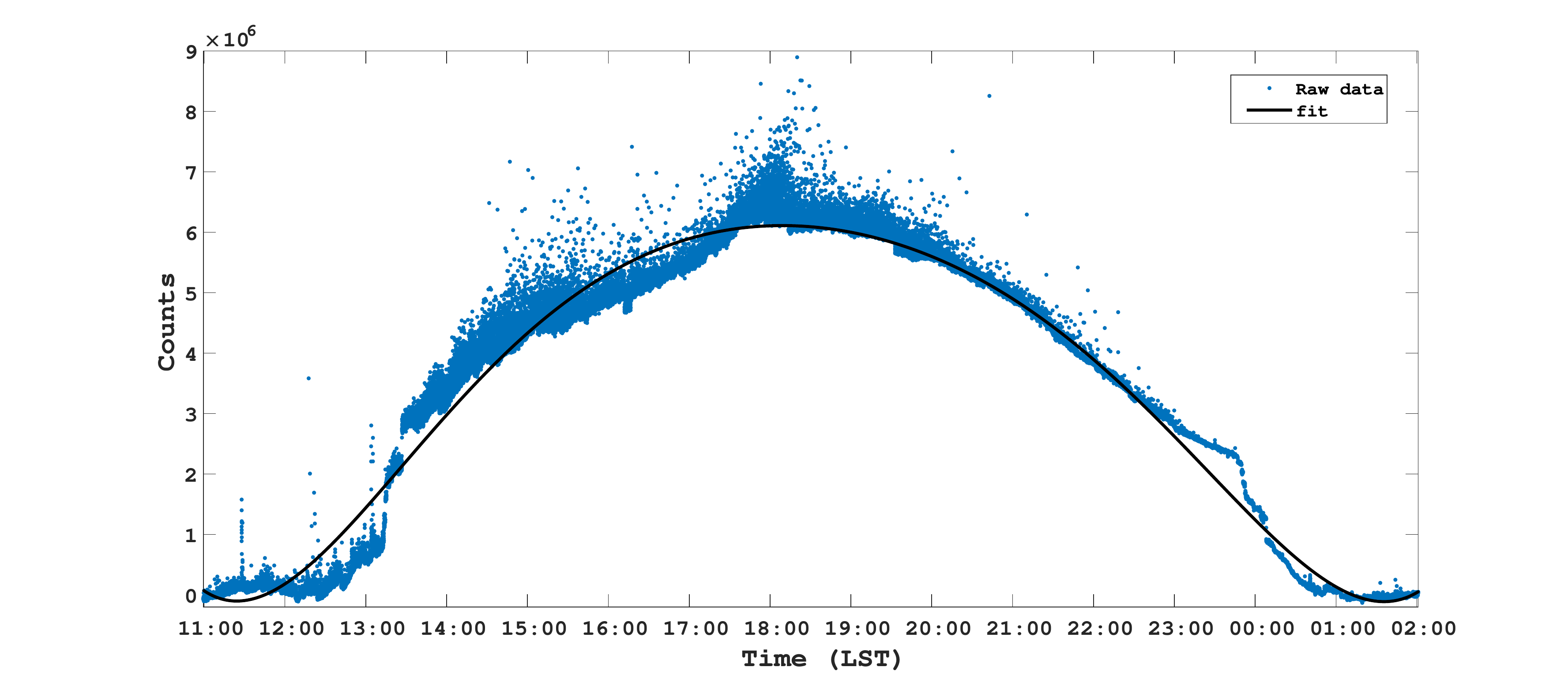}
\end{tabular}
\end{center}
\caption 
{\label{fig:figure7}
Observations of the sky background radiation with the pulsar array at 65\,MHz with an integration time of ${\approx}$\,1\,sec and bandwidth of ${\approx}$\,30\,MHz.  RFI was very minimal during the observing period.}
\end{figure} 

Moving on to pulsar observations, we targeted the historical B1919+21 (J1921+2153) for our trial run.
It is also located closer to the local zenith (${\approx}\,14^{\circ}$\,N) in Gauribidanur. Some of the characteristics of this pulsar 
are\cite{Bondonneau20}: mean flux at 65 MHz ${\approx}$\,1.6\,Jy, 
period\,=\,1.337\,sec, pulse width\,${\approx}$\,0.104\,sec, and 
DM\,${\approx}$\,12.437\,$\rm pc\,cm^{-3}$. Note that the dispersive nature of the ISM causes smearing of the pulsar signal over a time 
$\rm t_{smear}$ which depends on the DM of the pulsar, frequency and bandwidth of observation\cite{Issur02}. For the FWHM of the frequency bins mentioned earlier 
(${\approx}$\,36\,kHz) and the frequency band of observation in the present case
(i.e. 80\,-\,50\,MHz),
$\rm t_{smear}$  for B1919+21 will be in the range ${\approx}$\,5\,-\,34\,msec. These values are well within the pulse period of B1919+21 (i.e.\,1.337\,sec).  We observed the pulsar for a total duration of ${\approx}$\,2\,h with an integration time of ${\approx}$\,4\,msec. 
The north-south `beam' of the array was `phased' to the declination of the pulsar 
($21^{\circ}$\,N) for the observations (see Section 2).
Figure 8 shows the results of our observations. 
The pulsar was detected with S/N ${\approx}$\,23.
Comparing this with the mean flux of the pulsar, we find that the noise fluctuations should be ${\approx}$\,0.07\,Jy. This is reasonably consistent with the estimated 
$\rm{\Delta}S_{min}$ (${\approx}$\,0.12\,Jy) taking into consideration the 
above S/N and duty cycle (i.e. ratio of the width W of the pulse profile to its period P) of the pulsar to be ${\approx}$\,0.078\cite{Bondonneau20}. The other parameters used in the calculations are: total observing period 
${\approx}$\,2\,hr, $\rm T_{sys}\,{\approx}\,8500$\,K at 65 MHz\cite{Bondonneau20},
bandwidth ${\approx}$\,30\,MHz, and number of polarizations\,{=}\,1. The above observational results from Figures 7 \& 8
indicate that our observing system is well characterized. Note that for a total 
effective collecting area of $\rm {\approx}136\,m^{2}$ and an observing/integration period of 
${\approx}$1\,msec, the minimum detectable flux density of the system is ${\approx}$1\,kJy. This is lower than the peak flux density (${\sim}$100\,kJy) of the FRBs reported recently in the frequency range 400\,-\,800\,MHz\cite{Chime20}. Assuming that the flux densities of the FRBs are expected to increase towards the lower frequencies\cite{Keane16}, similar detections as well as observations of giant pulses from the Crab nebula pulsar\cite{Eftekhari16} are possible with our system.
We would like to add here that subsequent to the detection of B1919+21,   
the pulsars B0950+08, B0834+06, and B1133+16 were also successfully observed with our system. The related results will be reported separately. 

\begin{figure}[!t]
\begin{center}
\begin{tabular}{c}
\includegraphics[width=15cm]{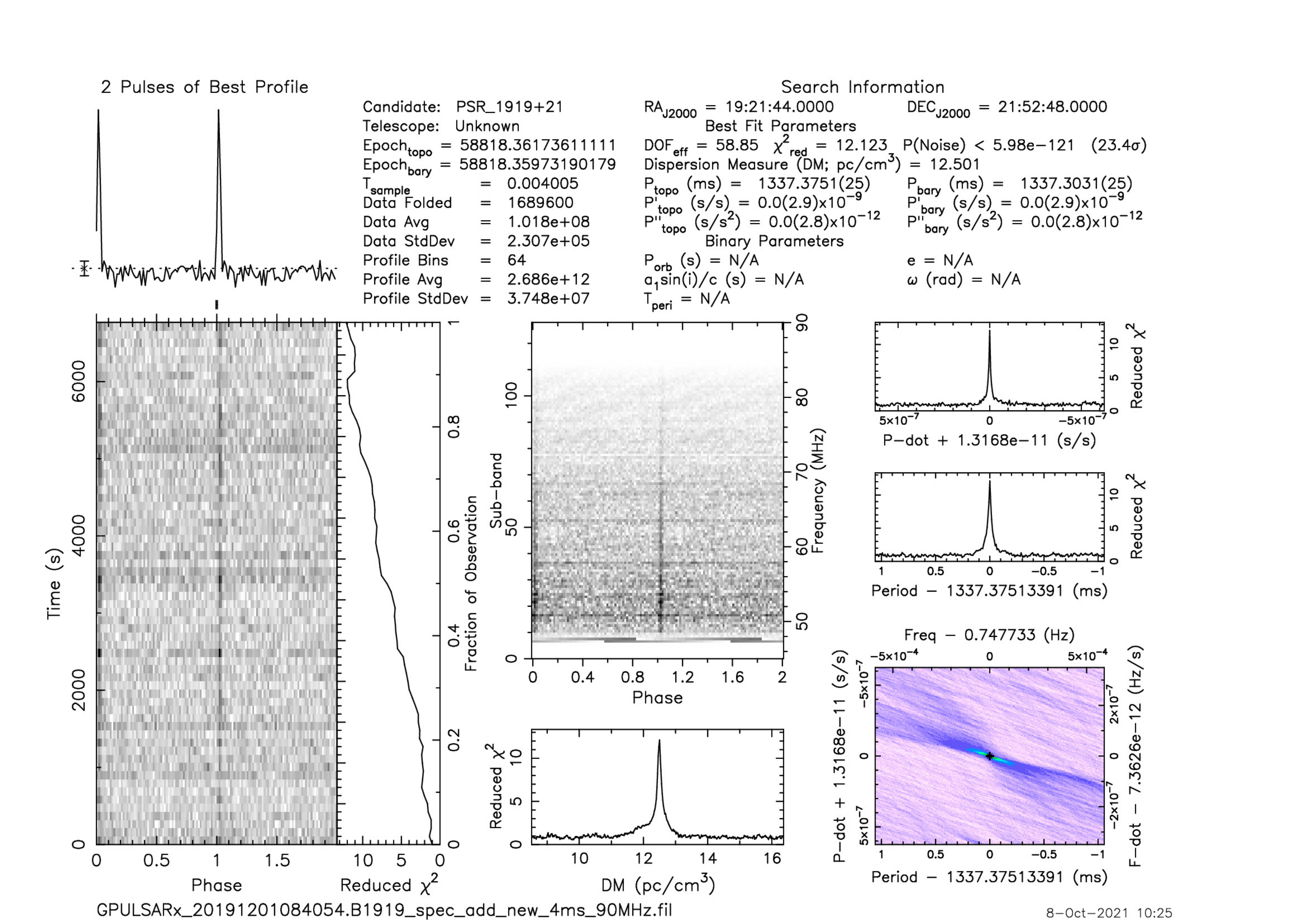}
\end{tabular}
\end{center}
\caption 
{\label{fig:figure8}
Observations of B1919+21 at 65\,MHz with the pulsar array in the Gauribidanur observatory. The data were analysed using PRESTO. RFI was very minimal during the observing period. They were mitigated using the `rfifind' tool in PRESTO (see Section 4).}
\end{figure} 


\section{Ongoing Work}
\label{sect:wip}

Further development is in progress to: (i) digitize the RF signal from individual antennas in the array. By applying different sets of delays while combining the signals, multiple beams can be formed to simultaneously observe different regions of the sky in declination. 
The array beam is already wide along RA (${\approx}110^{\circ}$). By having several beams in declination (each ${\approx}3^{\circ}$ wide), a larger area 
of the sky
can be monitored at the same time. This can be useful for observations of FRBs;
(ii) use Cross-polarized Log Periodic Dipoles (CLPD\cite{Sasi13}) and carry out polarization observations (see e.g. \citenum{Noutsos15}); (iii) use phase-coherent de-dispersion scheme with the DSPSR software package\cite{Bhat18}; (iv) increase the sampling frequency and thereby minimize the delay error relative to the source position (see Section 3); (v) use calibrated noise sources that can be switched into the signal path and/or routine observations of B1919+21 for calibration purposes. For polarization data, it is planned to use observations of radio transients from the Sun since some of them like type I bursts exhibit 100\% circular polarization\cite{McLean85}. 

\section{Summary}
\label{sect:sum}  

We have presented the characteristics of the  
low frequency (50\,-\,80\,MHz) observing system set up recently at the Gauribidanur observatory near Bangalore in India, and the initial observations that were carried out. 
Its large bandwidth (${\approx}$\,30\,MHz), 
wide sky coverage (${\approx}\,110^{\circ}{\times}80^{\circ}$), high spectral and temporal resolutions ($\rm {\approx}\,36\,kHz$ and 1\,msec, respectively), geographical 
location (${\approx}14^{\circ}$\,N), and the fact that the array is dedicated for time domain astronomy, are expected to be very useful to observe 
pulsars, radio bursts from the Sun\cite{Ramesh03,Mugundhan17}, and other high time resolution astrophysical phenomena like the FRBs.

\subsection*{Disclosures}
The authors declare that there are no conflicts of interest.

\subsection* {Acknowledgments}
We express our gratitude to the staff of the Gauribidanur observatory for their help in setting up the antenna/receiver systems, and carrying out the observations. We thank A. A. Deshpande for his encouragement and suggestions. We acknowledge the referees for their kind comments which helped us to present the results more clearly.

\subsection* {Data, Materials, and Code Availability} 
Data used in the study are available at https://www.iiap.res.in/gauribidanur/home.html.






\listoffigures

\end{spacing}
\end{document}